# Deterministic Design of Low-Density Parity-Check Codes for Binary Erasure Channels[1]

Hamid Saeedi and Amir H. Banihashemi

Department of Systems and Computer Engineering

Carleton University, Ottawa, Canada

{hsaeedi,ahashmei@sce.carleton.ca}

**Abstract- We propose a deterministic method to design irregular Low-Density Parity-Check (LDPC) codes for binary erasure channels (BEC). Compared to the existing methods, which are based on the application of asymptomatic analysis tools such as density evolution or Extrinsic Information Transfer (EXIT) charts in an optimization process, the proposed method is much simpler and faster. Through a number of examples, we demonstrate that the codes designed by the proposed method perform very closely to the best codes designed by optimization. An important property of the proposed designs is the flexibility to select the number of constituent variable node degrees *P*. The proposed designs include existing deterministic designs as a special case with *P = N*-1, where *N* is the maximum variable node degree. Compared to the existing deterministic designs, for a given rate and a given δ > 0, the designed ensembles can have a threshold in δ-neighborhood of the capacity upper bound with smaller values of *P* and *N*. They can also achieve the capacity of the BEC as *N*, and correspondingly *P* and the maximum check node degree tend to infinity.**

I. INTRDOUCTION

Low-Density Parity-Check (LDPC) codes have received much attention in the past decade due to their attractive performance/complexity tradeoff on a variety of communication channels. In particular, on the Binary Erasure Channel (BEC), they achieve the channel capacity asymptotically [1-4]. In [1],[5],[6] a complete mathematical analysis for the performance of LDPC codes over the BEC, both asymptotically and for finite block lengths, has been developed. For other types of channels such as the Binary Symmetric Channel (BSC) and the Binary Input Additive White Gaussian Noise (BIAWGN) channel, only asymptotic analysis is available [7]. For irregular LDPC codes, the problem of finding ensemble

---

degree distributions (denoted by $\rho(x)$ and $\lambda(x)$ for check nodes and variable nodes, respectively) that perform well (i.e., have the best threshold for a given rate or have the highest rate with negligible error or erasure probability for a given channel parameter) is called *code design*. For a variety of channels, the search for the best ensemble can be carried out based on different asymptotic analysis tools such as density evolution and Extrinsic Information Transfer (EXIT) charts [8-10] through an optimization process. In [1], a linear programming approach based on density evolution is used to find good degree distributions for the BEC. For the code design, there are two main categories in general: 1) For a given channel parameter, we look for a code with maximum rate and negligible probability of error or erasure; 2) For a given rate, the code capable of providing a reliable transmission for the worst possible channel parameter is designed. The second category is of more practical interest, while the first category is usually easier to design. For a given set of constituent variable and check node degrees, and for a given BEC parameter $\varepsilon$ (a given code rate $R$), the ensemble ($\rho(x),\lambda(x)$) which provides the highest reliable transmission rate (highest erasure protection) is called the *optimum ensemble.*

Optimization-based design methods are computationally expensive especially when a large number of constituent variable and check node degrees are permitted in the optimization process. In this paper, our aim is to deterministically design a close-to-optimum ensemble for a given check node degree distribution and a given number $P$ of constituent variable node degrees. The designed ensembles are expected to perform closely to the best ensembles designed by optimization. For both categories of code design, we consider two cases: A) The case where all the variable node degrees from 2 to a maximum degree $N$ are available ($P = N$-1); and B) the case where not all the degrees from 2 to $N$ are used ($P \neq N$-1). The ensembles designed in the two scenarios are referred to as Type-A and Type-B ensembles, respectively. In practice, the choice of $P$ may be affected by implementation considerations, where smaller values would be preferred. Although in this paper we focus on the design of ensembles for a given check node degree distribution, the designed ensembles can also be used to optimize both the variable node and the check node degree distributions iteratively in an optimization loop. In each iteration, $\rho(x)$ and subsequently $\lambda(x)$ (obtained by the method proposed in this paper), is modified to optimize the cost function (rate or threshold).

In [2-4], the authors introduce sequences of degree distributions that asymptotically achieve the capacity of a BEC for large values of maximum variable and check node degrees. For finite values of maximum variable and check node degrees, those sequences can also be used to deterministically design LDPC codes over a BEC. In fact, the constructions of [2-4] are a subset of constructions discussed in this paper (Type A in Category 2 of code design). Here, we show that more favorable solutions for finite values of $P$ do exist in our extended family of designs, i.e., for a given rate, a given check node degree distribution and a given $\delta > 0$, the designed ensemble can have a threshold in $\delta$-neighborhood of the capacity upper bound with a smaller value of $P$ and a smaller maximum variable node degree, compared to the ensembles of [2-4]. It should be noted that although the sequences of [2-4] are special cases of the designs proposed in this paper, the approach taken here to derive them is different and much simpler than that of [2-4]. In addition, it can be proved that by a proper choice of $P$ and for large values of $N$, the designed ensembles (in both categories and for both types) are capable of achieving the capacity of the BEC [11]. While this paper focuses on the code construction and simulation results for finite values of $P$ and $N$, asymptotic results on the constructed ensembles are presented in [11].

The paper is organized as follows. In the next section, we present a brief review of the BEC and its properties and define some notations that will be used throughout the paper. In section III, we discuss the first category of code design and prove a few lemmas that are used in the design process. Section IV generalizes the results of section III to the second category of code design. In section V, we provide some design examples. Section VI concludes the paper. The proofs of the lemmas, propositions and theorems are given in the appendix.

## II. PRELIMINARIES

We represent an LDPC code ensemble with its variable node and check node degree distributions:

$$\rho(x) = \sum_{i=2}^{D_c} \rho_i x^{i-1} \text{ and } \lambda(x) = \sum_{i=2}^{D_V} \lambda_i x^{i-1},$$

with constraints

$$\sum_{i=2}^{D_c} \rho_i = 1 \text{ and } \sum_{i=2}^{D_V} \lambda_i = 1, \tag{1}$$

where the coefficient of $x^i$ represents the percentage of edges connected to the nodes of degree $i+1$, and $D_v$ and $D_c$ represent the maximum variable node degree and the maximum check node degree, respectively. It should be noted that throughout the paper, we sometimes use $N$ to represent the maximum variable node degree. The difference between the two representations will be clear from the context.

Average check node and variable node degrees are given by

$$\bar{d}_c = 1/(\sum_{i=2}^{D_c} \rho_i / i) = 1/\int_0^1 \rho(x)dx \text{ and } \bar{d}_v = 1/(\sum_{i=2}^{D_V} \lambda_i / i) = 1/\int_0^1 \lambda(x)dx,$$

respectively.

For the code rate $R$, assuming a full-rank parity-check matrix, we have

$$R = 1 - \bar{d}_v / \bar{d}_c. \tag{2}$$

Consider a BEC with erasure probability $\varepsilon$. The capacity of this channel is $C = 1- \varepsilon$. For a given code ensemble over a BEC with a given channel parameter $\varepsilon$, the sufficient and necessary condition for the zero probability of message erasure after infinite number of iterations of a simple erasure recovery algorithm [1] is

$$\varepsilon\lambda(1-\rho(1-x)) < x \text{ for } 0 < x \leq \varepsilon \;.$$

This inequality can be rewritten as

$$\varepsilon\lambda(x) - 1 + \rho^{-1}(1-x) < 0, \; 0 < x \leq 1 \;. \qquad (3)$$

We call any code ensemble that satisfies (3) *convergent* for the given $\varepsilon$. For a code ensemble, the *threshold* is defined as the supremum of all $\varepsilon$ values that satisfy (3).

III. CODE DESIGN FOR THE HIGHEST RATE

In this section, we consider the case where we are given a check node degree distribution $\rho(x)$ and a certain channel erasure probability $\varepsilon$. Our goal is to find the variable node degree distribution $\lambda(x)$ of a convergent ensemble with the largest rate. If $N \; (\geq 3)$ denotes the maximum variable node degree, it is apparent from (2) that we need to minimize the average variable node degree or maximize its inverse:

$$\overline{d}_v^{-1} = \sum_{i=2}^{N} \lambda_i / i \;. \qquad (4)$$

In fact, the optimization of the rate is equivalent to maximizing $\overline{d}_v^{-1}$, subject to two constraints: equation (1) and inequality (3) which guaranties the code's convergence. From (4), it can be seen that in order to maximize $\overline{d}_v^{-1}$, higher percentages have to be assigned to lower degree variable nodes. The following lemma is a formulation of this idea.

*Lemma 1*: Consider a given check node degree distribution, a given channel parameter and a given set of constituent variable node degrees. Let $C$ be a convergent code ensemble of rate $R$ with variable node degree distribution $\lambda(x) = \sum_{i=2}^{N} \lambda_i x^{i-1}$. For given integer numbers $a$ and $b$ in the interval $[2, N]$, $a \neq b$, we form a new ensemble $C'$ with rate $R'$ such that $\lambda'_a = \lambda_a - k$, $\lambda'_b = \lambda_b + k$ and $\lambda'_i = \lambda_i$, for $i \notin \{a,b\}$ ($k$ is chosen such that $\lambda'_a \geq 0$ and $\lambda'_b \leq 1$). We then have:

1) If $a > b$, then $R' > R$.
2) If $a < b$, then $C'$ is convergent.

The first part of this lemma proposes a general approach to increase the rate but does not guarantee the convergence of the resulting ensemble. In fact, in conventional code optimization methods, the convergence of any newly constructed ensemble has to be verified by testing (3). In what follows, we derive upper bounds on $\lambda_i$ values based on the convergence condition (3) in the vicinity of $x = 0$. Then using these bounds, we can construct close-to-optimum ensembles whose convergence is *ensured* and need not to be checked by (3). To get such upper bounds, we consider the Taylor expansion of $\rho^{-1}(1-x)$. It can be shown that if $\rho(x)$ is a degree distribution, the Taylor series of $\rho^{-1}(1-x)$ around $x = 0$ is convergent [2]. Let

$$\rho^{-1}(1-x) = 1 - \sum_{i=2}^{\infty} T_i x^{i-1}, T_i > 0. \tag{5}$$

By replacing (5) in (3), we obtain

$$(\varepsilon\lambda_2 - T_2)x + (\varepsilon\lambda_3 - T_3)x^2 + ... + (\varepsilon\lambda_N - T_N)x^{N-1} - \sum_{i=N+1}^{\infty} T_i x^{i-1} < 0, \quad 0 < x \leq 1. \tag{6}$$

If $x$ tends to zero, all the terms with powers greater than one can be ignored compared to the first term on the left hand side of (6). Therefore, as $x$ tends to zero, we must have

.

$$(\varepsilon\lambda_2 - T_2)x \leq 0 \longrightarrow \lambda_2 \leq T_2/\varepsilon. \tag{7}$$

This is the upper bound on $\lambda_2$. Note that $T_2 = 1/\rho'(1)$ and thus (7) is the well-known stability condition $\varepsilon\lambda_2\rho'(1) \leq 1$ [8]. Now suppose that we set $\lambda_2$ equal to the upper bound of (7). Then, the first term on the left hand side of (6) becomes zero. In this case, as $x$ tends to 0, the term with $x^2$ becomes dominant and the necessary condition for convergence is

$$(\varepsilon\lambda_3 - T_3)x^2 \leq 0 \longrightarrow \lambda_3 \leq T_3/\varepsilon.$$

We can continue in a similar fashion and obtain an upper bound on $\lambda_i$, i.e., $\lambda_i \leq T_i/\varepsilon$, for $3 \leq i \leq N-1$, assuming that all $\lambda_j$ values for $j = 2,...,i$-1, have their maximum values.[2]

*1) Type-A*

Now assume that all variable node degrees from 2 to $N$ are available. The above inequalities suggest that for a given $\varepsilon$ and a given $\rho(x)$, the following ensemble, which is designed deterministically, could be a close-to-optimum candidate if it is convergent:

---

[2] It should be noted that this result coincides with the flatness condition proposed in [3] for capacity achieving sequences. The sequences of [3] however belong to the second category of code design.

$$\lambda_i = T_i / \varepsilon, \ 2 \leq i \leq N-1; \ \lambda_N = 1 - \sum_{i=2}^{N-1} \lambda_i . \tag{8}$$

We show that for a given $\rho(x)$ and a given $\varepsilon$, there exists a lower bound on $N$ that will ensure the convergence and an upper bound which guarantees $\lambda_N$ to be positive. The following lemma indicates that a unique $N$ satisfies both conditions. We call the corresponding degree distributions *Type-A*.

*Theorem 1*: Consider a given check node degree distribution $\rho(x)$, and denote the *i*th term of the Taylor expansion of $\rho^{-1}(1-x)$ at $x=0$ by $T_i$, as in (5). For a given channel parameter $\varepsilon \geq T_2$ and a set of constituent variable node degrees from 2 to $N$ ($N > 2$),[3] there exists a unique $N$ that satisfies the following bounds:

$$\sum_{i=2}^{N} T_i > \varepsilon, \tag{9}$$

$$\varepsilon \geq \sum_{i=2}^{N-1} T_i . \tag{10}$$

For such $N$, the convergence of Type-A ensemble is ensured and $\lambda_N \geq 0$.

Note that if we would like to design a code for a channel parameter $\varepsilon$ which is less than $T_2$, we have to decrease $T_2$ by increasing the average check node degree through the modification of $\rho(x)$.

*Theorem 2*: Consider the Type-A ensemble $C$ designed based on (8) for a given channel parameter $\varepsilon$. The channel parameter $\varepsilon$ is then the threshold of $C$.

We note that for check-regular codes with the check node degree $D_c$, there is a closed form expression for $T_i$ :

$$T_i = \binom{\alpha}{i-1}(-1)^i, \ \alpha = 1/(D_c - 1), \tag{11}$$

where $\binom{\alpha}{i}$ is defined as [2]:

---

[3] Note that based on (8), the condition $\varepsilon \geq T_2$ is equivalent to $\lambda_2 \leq 1$.

$$\binom{\alpha}{i} = \frac{\alpha(\alpha-1)...(\alpha-i+1)}{i!} = \frac{\alpha}{i}\left(1-\frac{\alpha}{i-1}\right)...\left(1-\frac{\alpha}{2}\right)(1-\alpha)(-1)^{i+1} . \qquad (12)$$

*Example 1*: For $\varepsilon = 0.48$ and $\rho(x) = x^5$, it can be seen that the value of $N$ which satisfies (9) and (10) is $N = 13$. The variable node degree distribution for Type-A ensemble is:

$\lambda(x) = 0.4167x + 0.1667x^2 + 0.1000x^3 + 0.0700x^4 + 0.0532x^5 + 0.0426x^6$
$+ 0.0353x^7 + 0.0300x^8 + 0.0260x^9 + 0.0229x^{10} + 0.0204x^{11} + 0.0165x^{12}$

This ensemble has rate $R = 0.4998$ and its threshold is 0.48.

Note that Type-A ensembles are optimal in a greedy sense, in that, starting from degree-2 variable nodes, we maximize the percentage of edges connected to lower degree variable nodes and thus aim for maximizing the rate of the ensemble. For a fixed check degree distribution and a given channel parameter, however, the value of $N$ and thus the number of constituent variable node degrees are both fixed and dictated by Theorem 1. In the following, we introduce new ensembles, where we have the flexibility to determine the number of constituent variable node degrees $P$ and design ensembles with $P < N-1$. The cost associated with reducing $P$ is a reduction in rate.

*2) Type-B*

Given a check node degree distribution and a channel parameter, Theorem 1 indicates that there exists a unique $N$ that satisfies (9) and (10). For such $N$, consider a variable node degree distribution which includes a few consecutive degrees starting from 2 and ending at $P<N$ and the maximum variable node degree $N$.

*Example 2*: Consider the Type-A ensemble of Example 1, in which $D_c = 6$, and the maximum variable node degree $N$ is 13. For $P = 4$, the new ensemble has variable node degrees 2, 3, 4 and 13.

For such constituent variable node degrees, a *Type-B* ensemble is constructed based on:

$$\lambda_i = T_i / \varepsilon, \ 2 \leq i \leq P; \ \lambda_N = 1 - \sum_{i=2}^{P} \lambda_i ,$$

where $T_i$ and $N$ are defined in (5) and Theorem 1, respectively.

Since $N$ satisfies the conditions of Theorem 1, part 2 of Lemma 1 will ensure the convergence of the new ensemble.

Consider now a Type-B ensemble with variable node degrees $2,3,...,P$, and a maximum variable node degree $N$. Based on part 1 of Lemma 1, if instead of $N$ we choose a smaller maximum variable node degree $D_v$ with the same percentage of adjacent edges, the newly constructed ensemble has a higher rate but can be non-convergent. By choosing the smallest $D_v$ which results in a convergent ensemble, we can create a new ensemble, referred to as *Modified Type-B* or *Type-MB*. Note that the variable node degree distribution for this ensemble is the same as that of Type-B ensemble with $\lambda_{D_v}$ replacing $\lambda_N$. Also, with an argument similar to that of Theorem 2, we can show that the thresholds of both Type-B and Type-MB ensembles are equal to the channel parameter.

*Example 3:* Consider the Type-B ensemble of Example 2 with $P = 4$. This ensemble has variable node degrees 2, 3, 4 and 13, with coefficients $\lambda_2 = 0.4167$, $\lambda_3 = 0.1667$, $\lambda_4 = 0.1000$ and $\lambda_{13} = 0.3176$, respectively. The rate of this ensemble is $R = 0.4679$, which is less than the rate of the Type-A ensemble of Example 1, as expected. If we keep decreasing the maximum variable node degree, we see that a degree distribution with degrees 2, 3, 4 and 8 is convergent while one with 2, 3, 4 and 7 is not. Therefore, for $\varepsilon = 0.48$ and $\rho(x) = x^5$, Type-MB ensemble has variable node degrees 2, 3, 4 and 8 with coefficients $\lambda_2 = 0.4167$, $\lambda_3 = 0.1667$, $\lambda_4 = 0.1000$ and $\lambda_8 = 0.3176$. This ensemble has a rate $R = 0.4926$ which is in between the rates of Type-A and Type-B ensembles, and in fact very close to the rate of Type-A ensemble. It is however important to note that compared to Type-A ensemble, which has 12 different variable node degrees with a maximum degree of 13, this ensemble has only 4 different variable node degrees and the maximum degree is only 8.

In the following proposition, we derive a lower bound on $D_v$ which is a sufficient condition for convergence.

*Proposition 1*: Consider an ensemble $C$ with a given check node degree distribution and a set of consecutive constituent variable node degrees from 2 to $P$ and a maximum variable node degree $D_v$ ($D_v \geq P+1$). Suppose that the channel parameter $\varepsilon$ is given and that $N$ is computed based on Theorem 1. Let $T_i$ be the $i_{th}$ term of the Taylor expansion of $\rho^{-1}(1-x)$, as in (5). For ensemble $C$, also let $\lambda_i = T_i/\varepsilon$, $2 \leq i \leq P$; and $\lambda_{D_v} = 1 - \sum_{i=2}^{P} \lambda_i$. Then the following lower bound on $D_v$ is a sufficient condition for the convergence of $C$:

$$D_v \geq N - \left(\sum_{i=P+1}^{N-1}(N-i)T_i\right)/(\varepsilon - \sum_{i=2}^{P} T_i) \ . \qquad (13)$$

For the ensemble of Example 3, the lower bound of (13) is equal to 7.8590, which suggests choosing $D_v = 8$. In this case, 8 is in fact the smallest possible value for $D_v$. In general however the lower bound of (13) may not result in the best possible answer for $D_v$. Nevertheless, one can use this lower bound as a starting point to conduct a quick search for the smallest $D_v$ which results in a convergent ensemble.

## IV. CODE DESIGN FOR THE HIGHEST THRESHOLD

In this section, we are interested in designing $\lambda(x)$ for an ensemble $C$ with a given check node degree distribution $\rho(x)$ and a certain rate $R$ that has the largest possible threshold. Suppose that the largest threshold is equal to $\varepsilon$, and is achieved by ensemble $C$. This implies that for the channel parameter $\varepsilon$, ensemble $C$ has the highest rate which is equal to $R$. This in turn suggests that a similar approach as the one described in Section III can also be applied to designing close-to-optimum ensembles for a given rate.

*1) Type-A*

For a given $\rho(x)$ and a given rate $R$, for Type-A ensembles, we consider the case where all variable node degrees from 2 to a maximum degree $N$ are available ($N$ will be determined later). Suppose that the threshold of the ensemble is equal to $\varepsilon$. We can then use (8) to compute $\lambda_i$ values based on $\varepsilon$. Using (2), we thus have

$$R = 1 - \frac{\overline{d}_c^{-1}}{1/N + \frac{1}{\varepsilon}\sum_{i=2}^{N-1} T_i(1/i - 1/N)}. \tag{14}$$

Solving this equation for $\varepsilon$ results in

$$\varepsilon = \frac{\sum_{i=2}^{N-1} T_i(1/i - 1/N)}{(1-R)^{-1}\overline{d}_c^{-1} - 1/N} = \frac{\sum_{i=2}^{N-1} T_i(1/i - 1/N)}{\overline{d}_v^{-1} - 1/N}. \tag{15}$$

The variable node degree distribution can then be computed as

$$\lambda_i = T_i \frac{(1-R)^{-1}\overline{d}_c^{-1} - 1/N}{\sum_{i=2}^{N-1} T_i(1/i - 1/N)}, \quad 2 < i \leq N-1; \lambda_N = 1 - \sum_{i=2}^{N-1} \lambda_i. \tag{16}$$

Now, for the ensemble to converge, it has to satisfy (9):

$$\sum_{i=2}^{N} T_i > \frac{\sum_{i=2}^{N-1} T_i(1/i - 1/N)}{\overline{d}_v^{-1} - 1/N}$$

and thus

$$\overline{d}_v^{-1} \sum_{i=2}^{N} T_i > \sum_{i=2}^{N} T_i/i. \tag{17}$$

Also, (10) should hold for the $N^{th}$ coefficient to be non-negative:

$$\overline{d}_v^{-1} \sum_{i=2}^{N-1} T_i \leq \sum_{i=2}^{N-1} T_i/i. \tag{18}$$

*Theorem 3*: For a given code rate $R$ and a given check node degree distribution (and thus a given $\bar{d}_c^{-1}$), if $R < 1 - 2/\bar{d}_c$,[4] then there exists a unique value of $N$ that satisfies (17) and (18).

Note that if the code rate $R$ does not satisfy the inequality of Theorem 3, we would have to increase $\bar{d}_c$ through modifying $\rho(x)$.

To summarize the design: For a given rate and a given check node degree distribution, we first find $T_i$ values, then compute $N$ from (17) and (18). Coefficients $\lambda_i$ are finally obtained based on (16). Note that with an argument similar to that of Theorem 2, one can show that the channel parameter obtained by (15) is in fact the true threshold of the Type-A ensemble.

*Example 4*: For $R = 0.5$ and $\rho(x) = x^5$, it can be seen that the value of $N$ which satisfies (17) and (18) is 13. The variable node degree distribution for the Type-A ensemble is

$\lambda(x) = 0.4169x + 0.1667x^2 + 0.1000x^3 + 0.0700x^4 + 0.0532x^5 + 0.0426x^6 +$
$0.0353x^7 + 0.0300x^8 + 0.0260x^9 + 0.0229x^{10} + 0.0204x^{11} + 0.0133x^{12}$.

This code has a threshold $\varepsilon = 0.4798$.

*2) Type-B*

Similar to Type-B ensembles of subsection III.2, in this subsection, we consider ensembles with a few consecutive variable node degrees from 2 to $P$ and a maximum degree $N$. We initiate the design by computing $N$ from (17) and (18) and then computing $\varepsilon$ from the following equation:

$$\varepsilon = \frac{\sum_{i=2}^{P} T_i (1/i - 1/N)}{(1-R)^{-1} \bar{d}_c^{-1} - 1/N} \quad . \tag{19}$$

We then set

$$\lambda_i = T_i / \varepsilon, \; 2 \leq i \leq P, \text{ and } \lambda_N = 1 - \sum_{i=2}^{P} \lambda_i . \tag{20}$$

*Theorem 4*: Coefficient $\lambda_N$ in (20) is positive.

In the following theorem, we also show that the new ensemble is convergent.

*Theorem 5*: A Type-B ensemble of rate $R$ is convergent over a channel with parameter equal to the value given in (19). Moreover this value is the threshold of Type-B ensemble.

---

[4] Note that this inequality is satisfied for any ensemble whose variable node degrees are at least two.

Similar to the case in subsection III.2, to obtain a better threshold for a given $P$, we can design a Type-MB ensemble by decreasing the maximum variable node degree from $N$ to a smaller value $D_v$. To design this ensemble, we use (19) and (20) with $N$ replaced by $D_v$, and find the smallest $D_v$ for which the ensemble is convergent for the channel parameter given by (19). With a similar argument used for Type-A and B ensembles, the value of (19) is the threshold of the Type-MB ensemble.

Unfortunately, for this category of code design, we have not been able to obtain a lower bound for $D_v$ similar to that of Proposition 1. One however can perform a maximum of $N-P$ trials to find the smallest $D_v$. Each trial consists of computing $\varepsilon$ from (19) and $\lambda(x)$ from (20), where in both equations $N$ is replaced by $D_v$. We then need to check whether inequality (3) holds for the tested $D_v$.

*Example 5:* Consider the Type-B ensemble with $P = 4$ and $D_c = 6$ corresponding to the Type-A ensemble of Example 4. This ensemble has variable node degrees 2, 3, 4 and 13, with coefficients $\lambda_2 = 0.4521$, $\lambda_3 = 0.1808$, $\lambda_4 = 0.1085$ and $\lambda_{13} = 0.2586$, respectively. The threshold of this ensemble computed by (19) is $\varepsilon = 0.4424$, which is less than that of the Type-A ensemble of Example 4, as expected. If we keep decreasing the maximum variable node degree, we see that a degree distribution with degrees 2, 3, 4 and 8 is convergent while one with 2, 3, 4 and 7 is not. Therefore, for $R = 0.5$ and $\rho(x) = x^5$, Type-MB ensemble has variable node degrees 2, 3, 4 and 8 with coefficients $\lambda_2 = 0.4266$, $\lambda_3 = 0.1706$, $\lambda_4 = 0.1024$ and $\lambda_8 = 0.3004$. This ensemble has a threshold $\varepsilon = 0.4688$, obtained from (19), which is in between the thresholds of Type-A and Type-B ensembles, and quite close to that of Type-A ensemble. It is however important to note that compared to Type-A ensemble, which has 12 different variable node degrees with a maximum degree of 13, this ensemble has only 4 different variable node degrees and the maximum degree is only 8.

## V. SIMULATION RESULTS

For simulation results, we consider both categories of code design for the highly popular check-regular ensembles[5]. We first consider an upper bound from [2] which is useful to measure the performance of our designed ensembles: For a given rate $R$ and a given average check node degree $\bar{d}_c$, the best achievable threshold is upper bounded by

$$\mathcal{Z} = (1-R)(1-R^{\bar{d}_c}).$$

By modifying the upper bound of [2], for a given channel parameter $\varepsilon$ and a given $\bar{d}_c$, we obtain the following upper bound on the best achievable rate:

$$\mathcal{R} = 1 - \frac{\varepsilon}{(1-(1-\varepsilon)^{\bar{d}_c})}.$$

---

[5] The popularity of check-regular ensembles is due to their better erasure correcting capabilities and simpler implementation in hardware. For a reference on the former, see [4].

*Example 6*: Consider the second category of code design for rate one half. Suppose that there is a constraint of $P = 4$ on the number of different variable node degrees. We consider check-regular ensembles with $D_c = 5$, 6, and 7. Table I shows the designed Type-MB ensembles for each check node degree. Note that for $R = 0.5$, the capacity upper bound implies $\varepsilon < 1-R = 0.5$. For each ensemble, we have shown the ratio of the threshold to $1-R = 0.5$ as well as to *3*.

As can be seen, the threshold improves by increasing $D_c$ and the ensemble with check node degree 7 achieves close to %97 of the upper bound *3*. In fact for $P = 4$, this is the best threshold than can be obtained by Type-MB check-regular ensembles. By increasing $D_c$ further, the threshold decreases unless we allow $P$ to also increase.

*Example 7:* In this example, by still focusing on check-regular ensembles, we allow $P$ to take values between 5 and 10 and for each value of $P$, we find the value of $D_c$ which results in the best threshold for Type-MB ensembles. These results are reported in Table II. As can be seen the threshold improves as $P$ is increased.

*Example 8:* In this example, we compare the performance of the Type-MB ensemble with $D_c = 7$ and $P = 8$ designed in Example 7 with its corresponding Type-A ensemble with the same check node degree distribution. The Type-A ensemble has the following variable node degree distribution:

$$\lambda(x) = .3394x + .1414x^2 + .0864x^3 + .0612x^4 + .0469x^5 + .0378x^6 + .0315x^7 + .0269x^8 + .0234x^9 + .0207x^{10}$$
$$+ .0185x^{11} + .0167x^{12} + .0152x^{13} + .0139x^{14} + .0128x^{15} + .0119x^{16} + .0111x^{17} + .0104x^{18} + .0097x^{19} + .0092x^{20}$$
$$+ .0087x^{21} + .0082x^{22} + .0078x^{23} + .0074x^{24} + .0071x^{25} + .0067x^{26} + .0065x^{27} + .0025x^{28}.$$

The threshold of this ensemble is equal to 0.4910 which is slightly better than that of Type-MB ensemble (0.4891). This is at the expense of 28 different variable node degrees (instead of 8 for Type-MB) and the maximum variable node degree of 29 (instead of 15 for Type-MB). In Fig. 1, we have shown the finite block length simulation results for the two codes. The block length is selected to be 5000 and the maximum number of iterations is limited to 200. For each simulation point, one hundred codeword erasures are generated. As can be seen, the two codes perform closely in finite block length as well.

In the next example, we demonstrate that for large enough maximum variable and check node degrees and $P$, Type-MB ensembles can practically achieve the capacity similar to the sequences of [2-4].

*Example 9:* We consider again rate one-half check-regular ensembles. With $D_c = 11$ and $P = 90$, we obtain $D_v = 203$ for a Type-MB ensemble. The threshold for this ensemble is equal to 0.4993 which is %99.9 of the capacity upper bound. If we use the check regular sequences of [4] with $D_c = 11$, to achieve the same percentage of the capacity bound, the designed code has to have 522 different constituent variable node degrees. Also, the maximum variable node degree increases from 203 to 523.

In the following, we compare our results with those obtained by optimization.

*Example 10:* From the database [12] of optimized LDPC codes, we consider the following check-regular rate one-half ($R = 0.5$) ensemble $C_1$ with $D_c = 7$:

$$\lambda_{C_1}(x) = 0.3354x + 0.1716x^2 + 0.0095x^3 + 0.0783x^4 + 0.1620x^5 + 0.1305x^{14} + 0.1126x^{15}.$$

For this ensemble, $P = 7$ and the threshold $\varepsilon_{C_1}$ is equal to 0.4917. For $P = 7$ and $D_c = 7$, the Type-MB ensemble $C_2$ has $D_v = 14$ with the following variable node degree distribution:

$$\lambda_{C_2}(x) = 0.3415x + 0.1423x^2 + 0.0870x^3 + 0.0616x^4 + 0.0472x^5 + 0.0380x^6 + 0.2824x^{13}.$$

This ensemble has a threshold $\varepsilon_{C_2} = 0.4880$ which is only slightly less than that of $C_1$. It should however be noted that the maximum variable node degree for $C_2$ (14) is smaller than that of $C_1$ (16). If we restrict the maximum variable node degree to 14 or smaller, we find another optimized ensemble $C_3$ in [12] with the following variable node degree distribution:

$$\lambda_{C_3}(x) = 0.2853x + 0.3135x^2 + 0.1162x^3 + 0.2848x^{12}.$$

Incidentally, this ensemble has the same threshold of 0.4880 as $C_2$ does. The advantage of this ensemble over $C_2$ is however the fewer number of constituent variable node degrees and the maximum variable node degree of 13 instead of 14.

*Example 11:* In this example, we consider the first category of code design. Let $D_c = 5$, $\varepsilon = 0.48$, and consider a constraint of $P = 4$ on the number of different variable node degrees. Using (9) and (10), we obtain $N = 7$. If we limit our search for an optimal ensemble to the maximum variable node degree of 7, there are a total of $\binom{6}{4} = 15$ different sets of constituent variable node degrees to be tested. We have performed optimization using exhaustive search for each set of constituent variable node degrees. Some of the best results are given in Table III. Note that for each row of Table III, the best $\lambda(x)$ which maximizes the rate for a convergent ensemble has been obtained. The highest ratio $R / \Re$ in the table is 0.9624, which corresponds to achieving about %96 of the rate upper bound.

Using our design method, we obtain a Type-MB ensemble with the following variable node degree distribution:

$$\lambda(x) = 0.5208x + 0.1953x^2 + 0.1139x^3 + 0.1699x^5.$$

The rate of this ensemble is 0.4769 which is more than %95 of the upperbound $\Re$. This is only slightly worse than the best result from the exhaustive search. The complexity of our design however is substantially lower than that of the exhaustive search.

VI. CONCLUSIONS

In this paper, we propose methods to deterministically design ensembles of irregular LDPC codes for binary erasure channels. The main idea is to maximize the percentage of edges connected to lower degree nodes in a greedy fashion. At finite maximum variable and check node degrees, the designed ensembles perform close to optimal, only slightly inferior to the ensembles designed by exhaustive search (or optimization algorithms with asymptotic analysis tools). This is while the design complexity of the proposed ensembles is substantially lower than those obtained by exhaustive search.

An important feature of the proposed design is the flexibility to choose the number of constituent variable node degrees $P$, and to deterministically design a close to optimal ensemble under this constraint. Check-regular ensembles designed by this method would provide attractive solutions for implementation.

The proposed ensembles are also capable of achieving the capacity as the maximum variable and check node degrees increase. While capacity-achieving ensembles of [4] are special cases of the proposed designs, the derivation approach in this paper is different and much simpler. Moreover, our extended family of designs include more attractive solutions for finite values of $P$, i.e., for a given rate, a given check node degree distribution and a given $\delta > 0$, the designed ensembles can have a threshold in $\delta$-neighborhood of the capacity upper bound with a smaller value of $P$ and a smaller maximum variable node degree, compared to the ensembles of [4].

A topic of future research would be to obtain similar results on deterministic designs of irregular LDPC code ensembles for other channels such as the AWGN channel.

## Appendix: Proof of the Lemmas, Propositions and Theorems

*Proof of Lemma 1:*

To prove claim 1, we show that $1/\bar{d}_v' > 1/\bar{d}_v$:

$$1/\bar{d}_v' = \sum_{i=2, i \neq a,b}^{D_V} \lambda_i / i + (\lambda_a - k)/a + (\lambda_b + k)/b > \sum_{i=2, i \neq a,b}^{D_V} \lambda_i / i + \lambda_a / a + \lambda_b / b = 1/\bar{d}_v,$$

where the inequality follows from $a > b$.

For claim 2, we show that the degree distribution for C' satisfies (3). For $0 < x \leq 1$, $a < b$, and $k \geq 0$, it is easy to see that $k.x^{b-1} - k.x^{a-1} \leq 0$. As a result, for $0 < x \leq 1$,

$$\lambda_a x^{a-1} + \lambda_b x^{b-1} \geq (\lambda_a - k)x^{a-1} + (\lambda_b + k)x^{b-1} = \lambda'_a x^{a-1} + \lambda'_b x^{b-1},$$

and therefore $\lambda(x) \geq \lambda'(x)$. This implies that if $\lambda(x)$ satisfies (3), so does $\lambda'(x)$, and thus C' is convergent.

## Proof of Theorem 1:

From (5), we have $\sum_{i=2}^{\infty} T_i = 1$, and since $T_2 \leq \varepsilon < 1$, and $T_i > 0, \forall i$, it is easy to see that there exists an integer $N$ that satisfies both (9) and (10) and that such $N$ is unique. For such $N$, to show the convergence of the Type-A ensemble, we verify (3). Substituting $\lambda_i = T_i / \varepsilon$, $2 \leq i \leq N-1$ in (3), we obtain:

$$(\varepsilon \lambda_N - T_N) x^{N-1} - \sum_{i=N+1}^{\infty} T_i x^{i-1} < 0, \ 0 < x \leq 1.$$

To show that the above inequality holds, we note that $\varepsilon \lambda_N = \varepsilon - \sum_{i=2}^{N-1} T_i$, and thus $\varepsilon \lambda_N - T_N = \varepsilon - \sum_{i=2}^{N} T_i < 0$, where the last inequality is the same as (9). Multiplying both sides by the positive value $x^{N-1}$, and subsequently subtracting the positive value of $\sum_{i=N+1}^{\infty} T_i x^{i-1}$ from the left hand side proves (3) and thus the convergence.

To prove that $\lambda_N$ is nonnegative, we divide both sides of (10) by $\varepsilon$, and obtain $\sum_{i=2}^{N-1} \lambda_i \leq 1$, which implies $\lambda_N \geq 0$.

## Proof of Theorem 2:

To prove that the channel parameter $\varepsilon$ is the threshold of the Type-A ensemble $C$, we show that if $C$ converges over a channel with parameter $\varepsilon'$, we must have $\varepsilon \geq \varepsilon'$. Based on (7), channel parameter $\varepsilon'$ must satisfy $\lambda_2 \leq T_2 / \varepsilon'$. By the construction of $C$, however, $\lambda_2 = T_2 / \varepsilon$, and therefore $T_2 / \varepsilon \leq T_2 / \varepsilon'$. This requires that $\varepsilon \geq \varepsilon'$.

## Proof of Proposition 1:

We denote the variable node degree distribution for ensemble $C$ with $\lambda(x)$. For the given check node degree distribution and channel parameter, we construct the Type-A ensemble $C'$ and denote its variable node degree distribution by $\lambda'(x)$. The maximum variable node degree for ensemble $C'$ is $N$. If $P = D_v - 1$, then $C$ will be the same as Type-A ensemble $C'$, and the proposition is trivial. We thus focus on $P < D_v - 1$.

First note that based on (3), we have

$$\varepsilon \lambda'(x) < 1 - \rho^{-1}(1-x), \ 0 < x \leq 1. \tag{A-1}$$

We define function $f$ as follows:

$$f(x) = \varepsilon\lambda'(x) - \varepsilon\lambda(x) = \sum_{i=2}^{N-1}T_i x^{i-1} + (\varepsilon - \sum_{i=2}^{N-1}T_i)x^{N-1} - \sum_{i=2}^{P}T_i x^{i-1} - (\varepsilon - \sum_{i=2}^{P}T_i)x^{Dv-1},$$

We therefore have

$$f(x) = \sum_{i=P+1}^{N-1}T_i x^{i-1} - (\varepsilon - \sum_{i=2}^{P}T_i)x^{Dv-1} + (\varepsilon - \sum_{i=2}^{N-1}T_i)x^{N-1}.$$

We now consider two cases: a) $D_v = N$ and b) $D_v < N$, and for both cases prove that $f(x)$ is strictly positive over the interval (0, 1). For case (a), combining the last two terms of $f(x)$ results in $(-\sum_{i=P+1}^{N-1}T_i)x^{N-1}$. It is then easy to see that $f(x)$ has only one sign change between consecutive nonzero coefficients. Based on Descartes' rule of signs, this implies that $f(x)$ has only one positive root. It is easy to see $f(1) = 0$, and thus $x = 1$ is the only positive root of $f(x)$. We also have $f(0) = 0$, and therefore $f(x)$ must be either strictly positive or strictly negative in the interval (0, 1). Using (13), it is a simple exercise to see $f'(1) < 0$, and thus $f(x)$ must be strictly positive over (0, 1). For case (b), $f(x)$ has two sign changes between consecutive nonzero coefficients. Based on Descartes' rule of signs, this implies that $f(x)$ has either zero or two positive roots. Since $f(1) = 0$, we must have the latter case. It is also easy to see that $f'(1) < 0$. This combined with $\lim_{x \to +\infty} f(x) = +\infty$ indicates that the other positive root of $f(x)$ must be larger than one, and thus $f(x)$ does not have any root in (0, 1). It therefore must be either strictly positive or strictly negative in the interval. Since $f'(1) < 0$, it must be the former.

Since $f(x)$ is strictly positive over (0, 1), we have $\varepsilon\lambda(x) < \varepsilon\lambda'(x)$ for $0<x<1$. This combined with (A-1) proves that ensemble $C$ is convergent.

*Proof of Theorem 3:*

Since $R > 0$, we have $\overline{d}_v < \overline{d}_c$, and thus

$$\int_0^1 \rho(x)dx < \overline{d}_v^{-1}.$$

Since for $0 < x < 1$, $0 < \rho(x) < 1$, one can easily see that $\int_0^1 \rho(x)dx = 1 - \int_0^1 \rho^{-1}(x)dx$. We therefore have

$$1 - \int_0^1 \rho^{-1}(1-x)dx = 1 - \int_0^1 \rho^{-1}(x)dx = \int_0^1 \rho(x)dx < \bar{d}_v^{-1} \ . \qquad \text{(A-2)}$$

Consider the following sequences for $n \geq 2$:

$$f(n) = \bar{d}_v^{-1} \sum_{i=2}^{n} T_i$$

$$g(n) = \sum_{i=2}^{n} T_i / i$$

To prove Theorem 3, we prove that there exists a unique $N$ that satisfies $f(N) > g(N)$ and $f(N-1) \leq g(N-1)$.

Both sequences $f$ and $g$ are strictly increasing. We note that based on inequality $R < 1 - 2/\bar{d}_c$, we have $\bar{d}_v^{-1} < 1/2$. Also, $f(2) = \bar{d}_v^{-1} T_2$ and $g(2) = T_2/2$. We thus have $f(2) < g(2)$. Moreover, we have

$$f(n) = \bar{d}_v^{-1} \sum_{i=2}^{n} T_i = \bar{d}_v^{-1} \sum_{i=2}^{n} T_i x^{i-1} \Big|_{x=1} \quad \text{and} \quad g(n) = \sum_{i=2}^{n} T_i / i = \sum_{i=2}^{n} (T_i x^i)/i \Big|_{x=1} \ .$$

Based on (5), this results in

$$\lim_{n \to \infty} f(n) = \bar{d}_v^{-1} (1 - \rho^{-1}(1-x)) \Big|_{x=1} = \bar{d}_v^{-1} ,$$

and

$$\lim_{n \to \infty} g(n) = \int_0^1 1 - \rho^{-1}(1-x)dx = 1 - \int_0^1 \rho^{-1}(1-x)dx < \bar{d}_v^{-1} ,$$

where the last inequality is based on (A-2). We thus have

$$\lim_{n \to \infty} f(n) > \lim_{n \to \infty} g(n) .$$

Putting this together with $f(2) < g(2)$, we conclude that there exist some finite $n>2$, for which $f(n) > g(n)$. Let $N$ be the smallest such $n$. We thus have

$$f(n) \leq g(n), \ 2 \leq n \leq N-1 . \qquad \text{(A-3)}$$

It is also easy to see that $N$ satisfies

$$\bar{d}_v^{-1} > 1/N , \qquad \text{(A-4)}$$

because otherwise, we have

$$\bar{d}_v^{-1} T_i \leq T_i / N \leq T_i / i, \ \text{for } 2 \leq i \leq N ,$$

and therefore

$$\bar{d}_v^{-1} \sum_{i=2}^{N} T_i \leq \sum_{i=2}^{N} T_i / i. \tag{A-5}$$

This means that $f(N) \leq g(N)$, which contradicts the above definition of $N$.

Now we prove that for any $n > N$, $f(n) > g(n)$. Using (A-4), we have

$$\bar{d}_v^{-1} T_i > T_i / N > T_i / i, \text{ for } i > N,$$

and thus for any $n > N$, we have

$$\bar{d}_v^{-1} \sum_{i=N+1}^{n} T_i > \sum_{i=N+1}^{n} T_i / i. \tag{A-6}$$

This together with $f(N) > g(N)$ results in

$$\bar{d}_v^{-1} \sum_{i=2}^{n} T_i > \sum_{i=2}^{n} T_i / i, \tag{A-7}$$

which is equivalent to $f(n) > g(n)$. This proves the uniqueness of $N$.

*Proof of Theorem 4*:

Note that based on the proof Theorem 3, the value of $N$ that satisfies (17) and (18) is unique and using (A-3), for any $P < N$, we have

$$\bar{d}_v^{-1} \sum_{i=2}^{P} T_i < \sum_{i=2}^{P} T_i / i. \tag{A-8}$$

Also,

$$\lambda_N = 1 - \sum_{2}^{P} \lambda_i = \frac{1}{\varepsilon}(\varepsilon - \sum_{2}^{P} T_i) = \frac{1}{\varepsilon}\left(\frac{\sum_{i=2}^{P} T_i(1/i - 1/N) - [(1-R)^{-1}\bar{d}_c^{-1} - 1/N]\sum_{2}^{P} T_i}{(1-R)^{-1}\bar{d}_c^{-1} - 1/N}\right) = \frac{1}{\varepsilon(\bar{d}_v^{-1} - 1/N)}\left(\sum_{i=2}^{P} T_i/i - \bar{d}_v^{-1}\sum_{i=2}^{P} T_i\right)$$

Based on (A-8) and (A-4), we conclude that $\lambda_N > 0$.

*Proof of Theorem 5*:

Let $C$ denote the Type-B ensemble. For the given check node degree distribution and code rate, let $C'$ denote the corresponding Type-A ensemble. Note that $C'$ is convergent on a channel with parameter equal to (15). Denote the threshold of ensemble $C'$ by $\varepsilon'$. We have

$$\lambda(x) = \sum_{i=2}^{P}(T_i/\varepsilon)x^{i-1} + (1 - \sum_{i=2}^{P} T_i/\varepsilon)x^{N-1},$$

and

$$\lambda'(x) = \sum_{i=2}^{N-1}(T_i/\varepsilon')x^{i-1} + (1-\sum_{i=2}^{N-1}T_i/\varepsilon')x^{N-1}.$$

Therefore

$$\varepsilon\lambda(x) = \sum_{i=2}^{P}T_i x^{i-1} + (\varepsilon - \sum_{i=2}^{P}T_i)x^{N-1}, \qquad (A\text{-}9)$$

and

$$\varepsilon'\lambda'(x) = \sum_{i=2}^{N-1}T_i x^{i-1} + (\varepsilon' - \sum_{i=2}^{N-1}T_i)x^{N-1}. \qquad (A\text{-}10)$$

Ensemble $C'$ is convergent by definition, therefore

$$\varepsilon'\lambda'(x) < 1 - \rho^{-1}(1-x), \text{ for } 0 < x \le 1. \qquad (A\text{-}11)$$

We also have

$$\varepsilon = \frac{\sum_{i=2}^{P}T_i(1/i - 1/N)}{(1-R)^{-1}\overline{d}_c^{-1} - 1/N} < \frac{\sum_{i=2}^{N-1}T_i(1/i - 1/N)}{(1-R)^{-1}\overline{d}_c^{-1} - 1/N} = \varepsilon',$$

since $P < N-1$ for Type-B ensemble, and $1/i > 1/N$ for $i \le N-1$. Therefore $\varepsilon < \varepsilon'$, and consequently

$$\sum_{i=2}^{N-1}T_i x^{i-1} + (\varepsilon - \sum_{i=2}^{N-1}T_i)x^{N-1} < \sum_{i=2}^{N-1}T_i x^{i-1} + (\varepsilon' - \sum_{i=2}^{N-1}T_i)x^{N-1}. \qquad (A\text{-}12)$$

Also, as $0 < x \le 1$, one can see that $T_i x^{i-1} - T_i x^{N-1} \ge 0$, for $P+1 \le i \le N-1$, and thus

$$\sum_{i=P+1}^{N-1}T_i x^{i-1} - \sum_{i=P+1}^{N-1}T_i x^{N-1} \ge 0. \qquad (A\text{-}13)$$

Adding (A-13) to (A-9), we obtain

$$\varepsilon\lambda(x) \le \sum_{i=2}^{N-1}T_i x^{i-1} + (\varepsilon - \sum_{i=2}^{N-1}T_i)x^{N-1}. \qquad (A\text{-}14)$$

Combining (A-14), (A-12) and (A-10), we have

$$\varepsilon\lambda(x) \le \sum_{i=2}^{N-1}T_i x^{i-1} + (\varepsilon - \sum_{i=2}^{N-1}T_i)x^{N-1} < \sum_{i=2}^{N-1}T_i x^{i-1} + (\varepsilon' - \sum_{i=2}^{N-1}T_i)x^{N-1} = \varepsilon'\lambda'(x),$$

and thus

$$\varepsilon\lambda(x) < \varepsilon'\lambda'(x). \qquad (A\text{-}15)$$

From (A-11) and (A-15) we conclude that

$$\varepsilon\lambda(x) < 1 - \rho^{-1}(1-x), \text{ for } 0 < x \le 1,$$

which proves the convergence. With an argument similar to that of Theorem 2, one can show that (19) is in fact the threshold of the Type-B ensemble.

**Table I: Type-MB ensembles for $R = 0.5$ and $P = 4$.**

| $D_c$ | $D_v$ | $\varepsilon_{MB}/0.5$ | $\varepsilon_{MB}/3$ | $\lambda(x)$ |
|---|---|---|---|---|
| 5 | 6 | 0.8873 | 0.9159 | $.5635x + .2113x^2 + .1233x^3 + .1019x^5$ |
| 6 | 8 | 0.9376 | 0.9525 | $.4266x + .1706x^2 + .1024x^3 + .3004x^7$ |
| 7 | 10 | 0.9610 | 0.9686 | $.3459x + .1445x^2 + .883x^3 + .4203x^9$ |

**Table II: Best Type-MB check-regular ensembles for $R = 0.5$ and different values of $P$.**

| $P$ | $D_c$ | $D_v$ | $\varepsilon_{MB}/0.5$ | $\varepsilon_{MB}/3$ | $\lambda(x)$ |
|---|---|---|---|---|---|
| 5 | 7 | 12 | 0.9624 | 0.9700 | $.3464x + .1443x^2 + .0882x^3 + .0625x^4 + .3586x^{11}$ |
| 6 | 7 | 13 | 0.9716 | 0.9793 | $.3431x + .1429x^2 + .0874x^3 + .0619x^4 + .0474x^5 + .3173x^{12}$ |
| 7 | 7 | 14 | 0.9761 | 0.9838 | $.3415x + .1423x^2 + .0870x^3 + .0616x^4 + .0472x^5 + .0380x^6 + .2824x^{13}$ |
| 8 | 7 | 15 | 0.9783 | 0.9860 | $.3407x + .1420x^2 + .0868x^3 + .0615x^4 + .0471x^5 + .0380x^6 + .0316x^7 + .2523x^{14}$ |
| 9 | 8 | 22 | 0.9836 | 0.9875 | $.2905x + .1245x^2 + .0771x^3 + .0550x^4 + .0425x^5 + .0344x^6 + .0288x^7 + .0247x^8 + .3225x^{21}$ |
| 10 | 8 | 23 | 0.9864 | 0.9902 | $.2897x + .1241x^2 + .0768x^3 + .0549x^4 + .0423x^5 + .0343x^6 + .0287x^7 + .0246x^8 + .0215x^9 + .3031x^{22}$ |

**Table III: Results for rate optimization of LDPC code ensembles with $D_c = 5$ and 4 different constituent variable node degrees at channel erasure probability 0.48.**

| Variable node degrees | $R$ | $R/я$ | $\lambda(x)$ |
|---|---|---|---|
| [2,3,4,6] | 0.4821 | 0.9622 | $.5206x + .0730x^2 + .4059x^3 + .0005x^5$ |
| [2,3,4,5] | 0.4822 | 0.9624 | $.5196x + .1172x^2 + .2950x^3 + 0.0682x^4$ |
| [2,4,5,6] | 0.4739 | 0.9455 | $.5208x + .4792x^3$ |

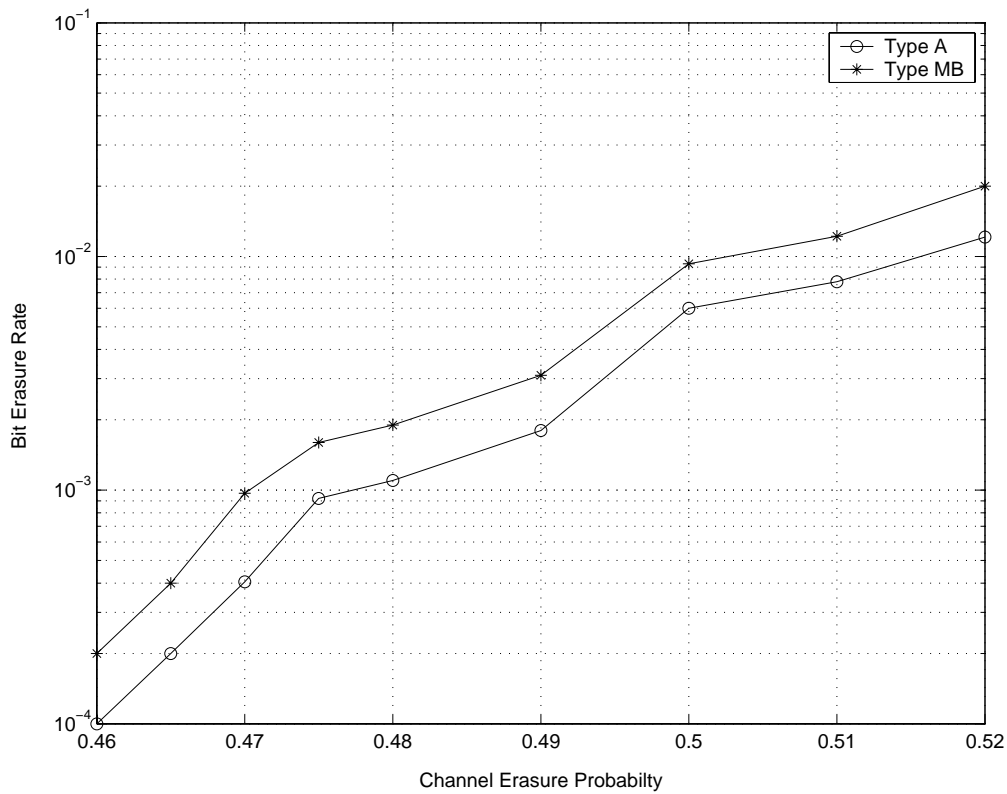

Figure 1: The performance of codes with block length 5000 from the ensembles of Example 8.